# Magnetic anisotropy in single crystal high entropy perovskite oxide La(Cr$_{0.2}$Mn$_{0.2}$Fe$_{0.2}$Co$_{0.2}$Ni$_{0.2}$)O$_3$ films

Yogesh Sharma[1], Qiang Zheng[1], Alessandro R. Mazza[1,2], Elizabeth Skoropata[1], Thomas Heitmann[3], Zheng Gai[4], Brianna Musico[5], Paul F. Miceli[2], Brian C. Sales[1], Veerle Keppens[5], Matthew Brahlek[1], and Thomas Z. Ward[1,*]

[1] *Materials Science and Technology Division, Oak Ridge National Laboratory, Oak Ridge, TN 37831, USA*
[2] *Department of Physics and Astronomy, University of Missouri, Columbia, MO 65211, USA*
[3] *University of Missouri Research Reactor, University of Missouri, Columbia, MO 65211, USA*
[4] *Center for Nanophase Materials Science, Oak Ridge National Laboratory, Oak Ridge, TN 37831, USA*
[5] *Department of Materials Science and Engineering, The University of Tennessee, Knoxville, TN 37996, USA*

*\* wardtz@ornl.gov*

## ABSTRACT

Local configurational disorder can have a dominating role in the formation of macroscopic functional responses in strongly correlated materials. Here, we use entropy-stabilization synthesis to create single crystal epitaxial ABO$_3$ perovskite thin films with equal atomic concentration of *3d* transition metal cations on the B-site sublattice. X-ray diffraction, atomic force microscopy, and scanning transmission electron microscopy of La(Cr$_{0.2}$Mn$_{0.2}$Fe$_{0.2}$Co$_{0.2}$Ni$_{0.2}$)O$_3$ (L5BO) films demonstrate excellent crystallinity, smooth film surfaces, and uniform mixing of the *3d* transition metal cations throughout the B-site sublattice. The magnetic properties are strongly dependent on substrate-induced lattice anisotropy and suggest the presence of long-range magnetic order in these exceptionally disordered materials. The ability to populate multiple elements onto a single sublattice in complex crystal structures opens new possibilities to design functionality in correlated systems and enable novel fundamental studies seeking to understand how diverse local bonding environments can work to generate macroscopic responses, such as those driven by electron-phonon channels and complex exchange interaction pathways.

## I. INTRODUCTION

$ABO_3$ perovskite transition metal oxides have a wide range of functionalities and often possess strong electron correlations driven by the material's specific cation composition. The manipulation and control of the cations' valence, ionic radius, spin composition, *etc* is then of central interest to both fundamental and applied studies. Substitutional doping is a cornerstone of experimental approaches used to explore mechanistic responses to changes in the charge, orbital, lattice, and spin contributions of these strongly correlated systems. The nearly degenerate energy scales of these parameters also mean that the minute extrinsic shifts to untargeted values, as with ion size variance, must also be considered. In correlated materials, even well diluted disordering from substitutional doping can play an important or even dominating role in the emergence of macroscopic behaviors, such as, high $T_c$ superconductivity[1,2], magnetic texture, scalable non-fermi liquid responses[3,4], and metal-insulator transitions[5,6]. Direct studies on the role of disorder are, however, difficult. First principle methods that rely on periodic boundary conditions fundamentally fail to capture the non-periodic nature of disorder, and approximation, such as large unit cells or effective disorder potentials are either not practical or of limited accuracy. Further, experimentally synthesizing crystalline systems with high levels of uniformly disperse disorder is exceptionally challenging, as the traditional substitutional doping route is limited by the constraints of thermodynamic formation energies[7]. Increasing substitution on a single site beyond a few percent often results in quenching of the uniform disorder in the form of extrinsic secondary structural phase formation or compensating vacancy creation. Recent works utilizing entropy stabilization during synthesis demonstrates an important new route to bypass these issues and allow access to the strong-limit of disorder.[8,9]

Entropy's role in synthesis has received increased attention over the past decade with the development of the field of high entropy alloys (HEA).[10] By combining 5 or more elements in roughly equiatomic ratios, alloys of exceptional structural properties have been discovered. These works have focused predominantly on combining transition metals onto a single lattice. As a result, most novel properties accessible to development tend to be limited to those of a mechanical nature, such as high melting temperatures, excellent radiation tolerances, and high tensile strengths. The focus on single-sublattice metallic alloys has meant that the mixing and stabilization mechanism is still contested. The introduction of high entropy oxides (HEO) recently changed this discussion[8] by suggesting that the addition of a single-element anion sublattice to the multicomponent cation sublattice maximizes entropy in the crystal. In effect, electrons in single-sublattice metallic HEA are non-localized, which can act to quench disorder and lower the number of possible microstates. The addition of an intermediary anion sublattice introduces localization, which increases the number of microstates available to the macroscopic system. As an example, by introducing localization, the orbital overlap can be reduced, which increases the importance of double and super exchange. Similarly, increasing localization can be used to bring the kinetic energy scale to a similar level as the onsite Coulomb potential; this is significantly different than the case of the highly delocalized electrons in metallic HEAs where the screening of the Coulomb potential is large so kinetic terms dominate, and direct exchange dictates the magnetism. While initial HEO work focused on the addition of a single anion sublattice to a single cation sublattice in the form of a rocksalt[8,11] lattice, many other structures are currently under development. These include forms in which two or more cation sublattices coexist with the uniform anion sublattice, such as spinels[12] and perovskites populated by up to 10 cations[9,13]. This ability to create such a wide and varied range of new HEO's promises to open new opportunities in functional materials similarly to the discoveries made in HEA's for structural materials[14,15].

In this work, the first examples of single crystal epitaxially strained La(Cr$_{0.2}$Mn$_{0.2}$Fe$_{0.2}$Co$_{0.2}$Ni$_{0.2}$)O$_3$ films are presented. X-ray diffraction, atomic force microscopy, and scanning transmission electron microscopy demonstrate excellent crystallinity, smooth film surfaces, and uniform mixing of the *3d* transition metal cations throughout the B-site sublattice. Field dependent magnetometry shows that these materials have strong magnetic anisotropy and may contain long-range magnetic order.

## II. EXPERIMENTAL DETAILS

A ceramic target of stoichiometric La(Cr$_{0.2}$Mn$_{0.2}$Fe$_{0.2}$Co$_{0.2}$Ni$_{0.2}$)O$_3$ (L5BO hereafter) was synthesized using the conventional solid-state reaction method [16]. Pulsed laser epitaxy was used to grow L5BO thin films on the three different substrates, including LaAlO$_3$ (LAO, $a$ = 3.788 Å), (LaAlO$_3$)$_{0.3}$(SrAl$_{0.5}$Ta$_{0.5}$O$_3$)$_{0.7}$ (LSAT, $a$ = 3.868 Å), and SrTiO$_3$ (STO, $a$ = 3.905 Å). A KrF excimer laser ($\lambda$ = 248 nm) operating at 5 Hz was used for target ablation. The laser fluence was 0.85 J/cm$^2$ with an area of 3.5 mm$^2$ on the target. The target-substrate distance was set at 5 cm. Deposition optimization was performed and the optimal growth conditions were found to occur with an oxygen partial pressure of 90 mTorr at a substrate temperature of 625 °C. Thin film samples (48 nm in thickness) of each strain state were grown. After deposition, the films were cooled to room temperature under 100 Torr oxygen pressure.

The crystal structure and growth orientation of the films were characterized by X-ray diffraction (XRD) using a four-circle high resolution X-ray diffractometer (X'Pert Pro, Panalytical) (Cu K$\alpha_1$ radiation). Atomic force microscopy (Nanoscope III AFM) was used in tapping mode to monitor surface morphology of the as-grown films with all films showing <1 nm rms surface roughness. Cross-sectional specimens oriented along the [100] STO direction for scanning transmission electron microscopic (STEM) analysis were prepared using ion milling after mechanical thinning and precision polishing. High-angle annular dark-field (HAADF) and electron-energy loss spectroscopy (EELS) analysis were carried out in a Nion UltraSTEM100, equipped with a cold field-emission electron source and a corrector of third- and fifth-order aberrations, operated at an accelerating voltage of 100 kV. A probe convergence angle of 30 mrad and an inner detector angle of about 86 mrad was used for HAADF imaging. EELS analysis was carried out using a Gatan Enfinium spectrometer with a collection semiangle of 48 mrad. Magnetization as a function of temperature and applied field, M(T) and M(H), respectively, were recorded using a Quantum Design MPMS3 superconducting quantum interference device (SQUID) magnetometer.

### III. RESULTS AND DISCUSSION

Figures 1(a) and (b) show θ-2θ XRD scans for L5BO films deposited on LAO, LSAT, and STO substrates. XRD scans demonstrate that all films on the three different substrates are single-crystal c-axis-oriented epitaxial films with well-defined Laue fringes, indicating homogeneous atomically flat surfaces and interfaces of the epitaxial thin films. The in-plane epitaxial orientation relationship of the films to the substrates is cube-on-cube to the (001)-oriented LAO, LSAT, STO substrates: (001) L5BO ∥ (001) LAO/LSAT/STO; [100] ∥ [100]. Rocking curves on the (002) peak in ω show full width at half maximum (FWHM) of 0.11°, 0.08°, 0.09° for films on LAO, LSAT, and STO, respectively (see supplemental). Figure 1(c) shows reciprocal space maps (RSMs) around the (103) peaks for the L5BO films and substrates. The films grown on STO and LSAT are coherently strained. The film grown on LAO shows some relaxation away from the substrate at the 48 nm thicknesses presented here. Using the bulk lattice parameter found in the ceramic L5BO target of 3.871Å, the applied strain mismatches are given as -2.2% for LAO, -0.08% for LSAT, and +0.87% for STO. Atomic force microscopy shows the surface morphologies of these films to present clean step and terrace structures with atomically smooth surfaces (Fig. 2).

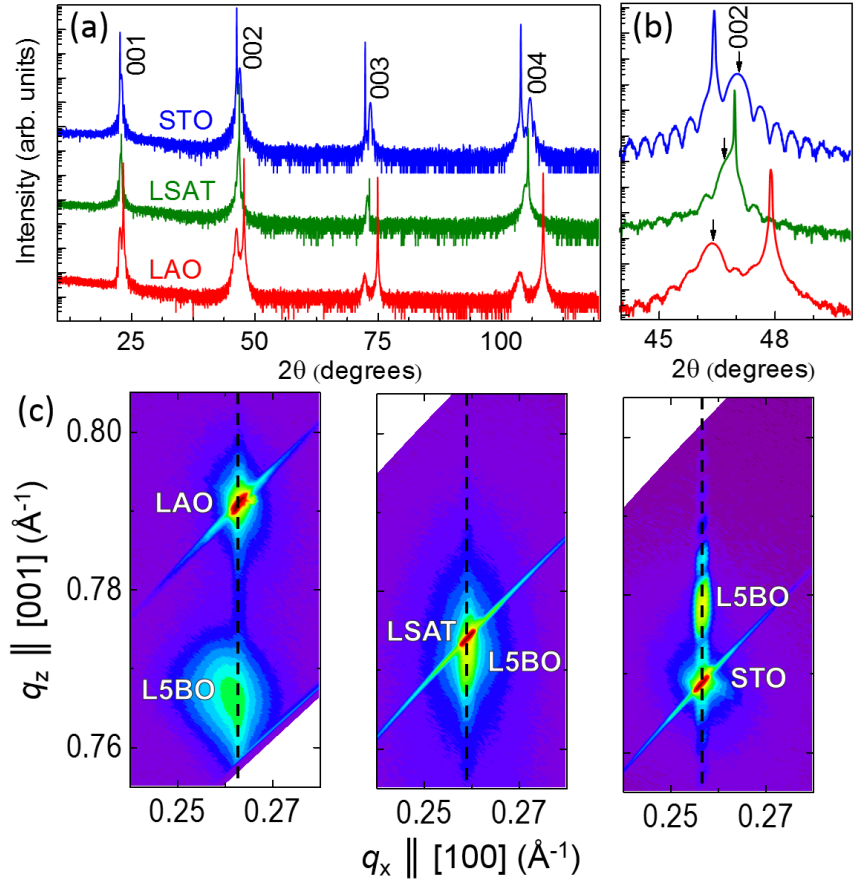

**FIG. 1.** Epitaxial growth of highly disordered perovskite films: (a) X-ray θ−2θ diffraction patterns of the La(Cr$_{0.2}$Mn$_{0.2}$Fe$_{0.2}$Co$_{0.2}$Ni$_{0.2}$)O$_3$ [L5BO] films (48 nm) grown on STO, LSAT, and LAO substrates. (b) θ−2θ diffraction patterns around the substrate 002 peaks, where the film peaks and substrates are denoted as arrows and asterisks, respectively. All L5BO films are single-crystal c-axis-oriented epitaxial films with well-defined Laue fringes. A systematic decrease in the c-axis lattice constants of the films is observed with changing strain-state from compressive (on LAO) to tensile (on STO). (c) Reciprocal space mapping around the 103 reflections of 48 nm L5BO films and substrates.

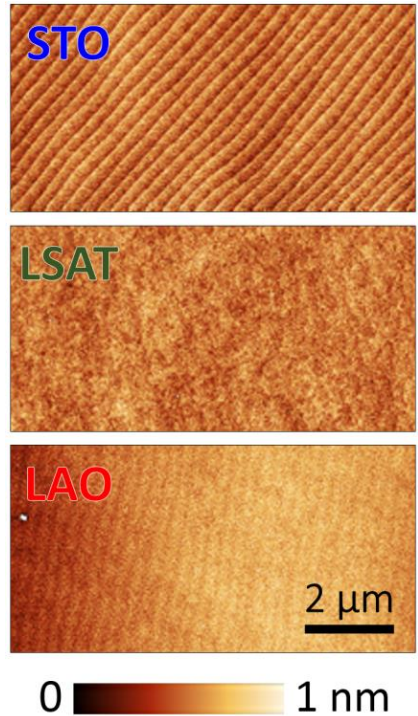

**FIG. 2.** Atomic force microscopy shows atomically smooth surfaces with step and terrace structures for films grown on each of the three substrates.

The microstructure and atomic resolution compositional mapping of L5BO films grown on STO are examined by STEM-EELS. The microstructure of the L5BO/STO interface shows a single crystalline lattice with an abrupt and fully coherent interface with the substrate.[16] The STEM observations are consistent with the XRD findings that the films are uniform and epitaxial. Figure 3(a) shows the HAADF image and the area analyzed by EELS (marked by white dashed-box). The atomically resolved maps of Cr-L, Mn-L, Fe-L, Co-L, and La-M signals are shown in Fig. 3(b). We see that La is only observed on the A-site sublattice while Cr, Mn, Fe, and Co are uniformly and randomly distributed on the B-site sublattice. Each B-site column contains each of the transition metal cations. We note that the Ni-L edge could not be cleanly resolved due to its proximity to La-M. This is a well-known issue for resolving Ni. The lack of observable "dark" pockets in EELs, lack of visible secondary phases in XRD, and preference of Ni toward B-site coordination can be used to rule out local Ni clustering, parasitic secondary Ni phase formation, and A-site mixing (see supplemental). This implies that Ni is also uniformly distributed similarly to the other transition metal cations on the B-site sublattice. The combination of x-ray diffraction with local electron microscopy demonstrates that the films are of excellent crystallinity and possess fully atomically disordered and mixed cations in the quinary B-site sublattice. It is worth noting that the synthesis of the presented films took very little growth optimization, and while a full growth diagram is outside the scope of this work, the authors would like to anecdotally report that from their experience the synthesis of these materials is far simpler than many other perovskite oxides containing fewer cations.

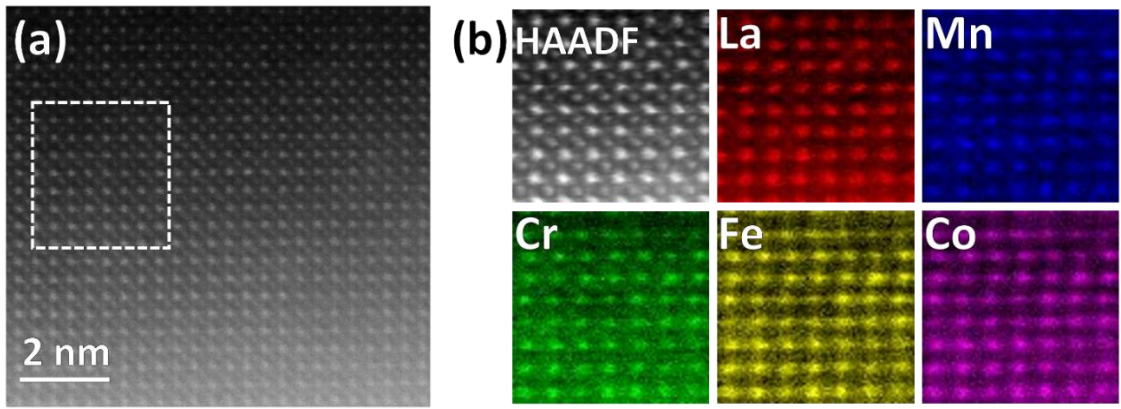

**FIG. 3.** STEM imaging and chemical mapping. (a) Magnified cross-sectional HAADF-STEM image for a L5BO film grown on STO, along the [100] STO direction. (b) HAADF survey image marked by white-dashed box in (a), where EELS mapping was performed with EELS elemental maps of La-*M*, Mn-*L*, Cr-*L*, Fe-*L*, and Co-*L* EELS signal from the region. The EELS mapping shows homogeneous distribution of B-site cations and thereby confirm uniform disorder at atomic scale.

To understand how this exceptional level of disorder impacts functionality, the electrical and magnetic properties are examined. All films are insulating and show no appreciable electrical response to the application of high magnetic fields nor a significant response to substrate induced strain.[16,17] This is not the case with magnetic properties. SQUID magnetometry is used to probe magnetization as a function of magnetic field applied in-plane and out-of-plane. Figure 4 shows the different magnetization loops for each of the differently strained films. In the tensile strained film grown on STO with magnetic field directed in-plane, the magnetization shows very little hysteresis and a relatively weak saturation moment ranging from ~0.12 $\mu_B$/uc to ~0.23 $\mu_B$/uc as temperature is reduced from 10 K to 2 K. Changing field direction to out-of-plane shows a dramatic difference in magnetic response. There is a hardening of magnetization over all, but most interesting is the development of a temperature dependent hysteresis. As temperature drops from 10 K to 2 K, the loop shape evolves from an elongated squared loop to a wasp-waisted multi-phase loop, whose relatively low magnetization suggests that the film possess a smaller ratio of ferromagnetism with a larger portion being antiferromagnetically aligned. The presence of multiple magnetic phases was recently reported in ceramic[9] L5BO, and this single crystal behavior confirms that this property is inherent to the material. However, the coherently strained applied on the single crystal films show a striking difference when compared to the ceramic; the application of the magnetic field along the in-plane direction removes the antiferromagnetic signature and shows a clean soft magnetic behavior with a very low coercive field. This field direction dependence is reversed in the compressively strained films. For the film grown on LSAT, the easy axis of magnetization is in the out-of-plane direction and the harder squared loop is now observed in the in-plane direction. The apparent reduction in feature size is the result of an overall increase in magnetization which acts to somewhat obscure the signature of the smaller secondary phase; this may be of some significance in that it may signal that the absolute amount of material hosting this magnetic phase is the same on both the tensile and compressively strained films and would suggest that surface symmetry breaking may be significant. Finally, for the most strongly compressively strained and partially relaxed film grown on LAO, there is a strong increase in magnetization to 1.5 µB/uc. At the lowest temperatures, this increase appears to swamp the observable hysteresis which is still seen on the loop taken at 10 K with field applied in-plane.

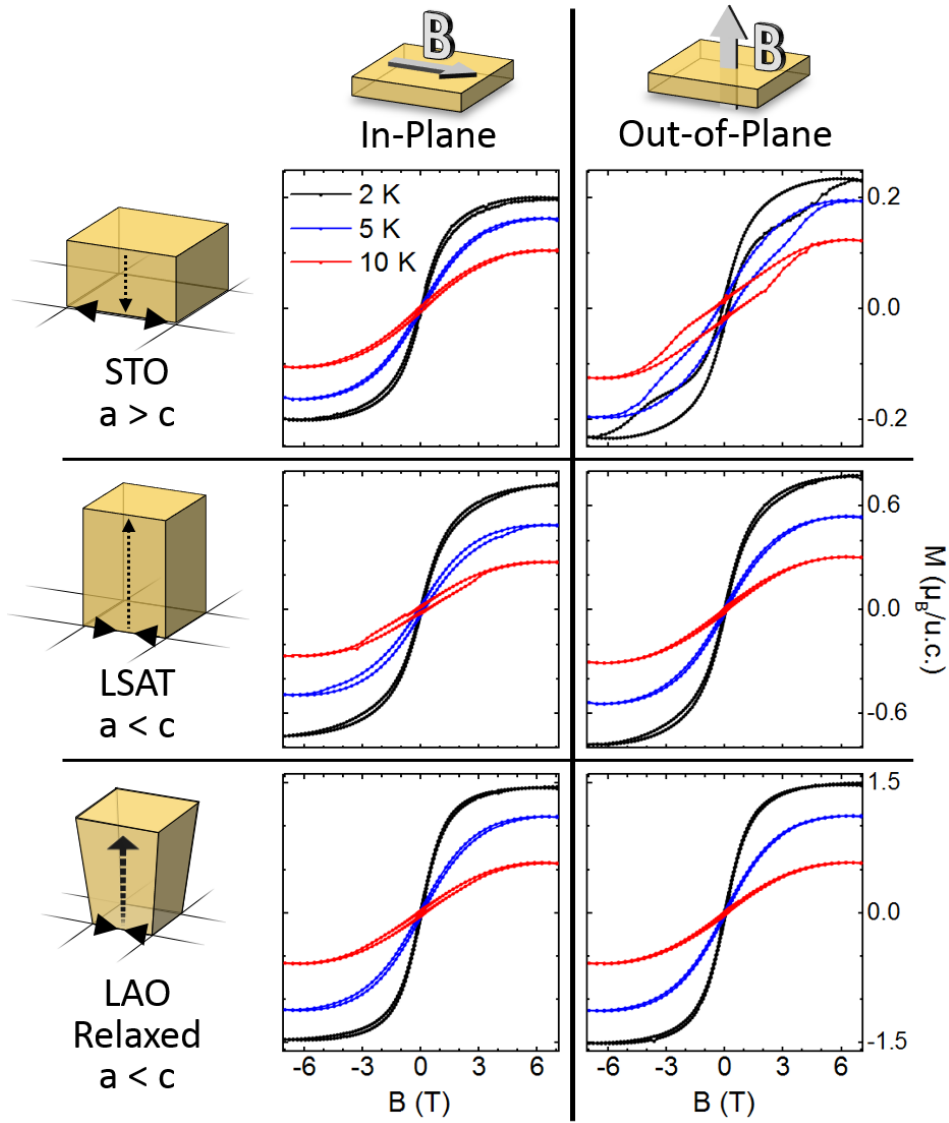

**FIG. 4.** The role of strain anisotropy on magnetization. Left (Right) column shows magnetization loops at three different temperatures with field directed in-plane (out-of-plane). In diagrams, notation on lattice is given *a* and *c* as in-plane and out-of-plane lattice constants, solid lines along in-plane direction denote epitaxy-induced deformation direction on the film while dotted lines represent Poisson-driven response to the underlying substrate strain. Magnetization as a function of magnetic field on the differently strained films shows that the directions of lattice tetragonality has a strong influence on the magnetic easy axis direction and absolute magnetization.

An important observation from the magnetization responses is that the saturation moment appears to be highly dependent on lattice symmetry. Consider that the reported saturation moment at 10 K for the fully relaxed ceramic was recently reported as 0.2 µB/uc, which follows the trend observed in this work—the +0.87% strained film has a moment of 0.10 µB/uc, the -0.08% strained film has marginally higher moment of 0.28 µB/uc, and significantly lower than the -2.2% strained sample's moment of 0.58 µB/uc. This observation can be explained if the modification of the superexchange coupling between the transition metal (TM) atoms is of central importance in driving the observed anisotropy. Specifically, the compression of the in-plane *a* and *b* lattice parameters and elongation of the *c* axis in the samples grown on LSAT and LAO would drive an increase in the covalent overlap on the TM-O-TM bonds in both the *a* and *b* direction in the plane and a reduction of overlap along the apical *c* direction. In the tensile strained film on STO, the TM-O-TM bond overlap is reduced along both in-plane directions and increased along the apical *c* direction. This increase in orbital overlap and crystal field could then be the driver of the enhanced ferromagnetic signature and the higher saturation moment. To clarify this point, further measurements aimed at quantifying element specific charge state and moment will be needed. Additionally, to answer the question of surface vs bulk magnetic states a technique such as polarized neutron reflectometry should be considered.

Studying the impact of high configurational disorder in cation distribution will allow a deeper insight into how disorder impacts local order parameter coupling and gives rise to the emergence macroscopic behaviors[9,11,14,18,19]. Modification of the A and B sites through substitutional doping allows for a wide variety of distortions and charge states to be tuned[20–22]. Structure and order parameters are strongly affected by the relationships of cations' oxidation state to cage distortion (e.g., Jahn-Teller distortions and octahedral tilts and rotations) [23–25]. In correlated materials, these shifts to local environment are often central to the emergence of important macroscopic functionalities[24,26,27]. Such an ability to use entropy stabilization to synthesize perovskite oxides with extreme configurational disorder, opens new possibilities for designing materials that show unusual sensitivity to disorder. The entropy stabilization approach provides an ideal opportunity to stabilize single phase highly disordered oxides with 5 or more elements on a single sublattice[8,9]. In these materials, entropy acts to enhance mixing and single phase formation as the random distribution of constituent elements into the cation sublattice(s) enhances the configurational entropy in such oxide solutions [8,9,14,28–31].

## V. CONCLUSION

Single crystal epitaxial thin films of $La(Cr_{0.2}Mn_{0.2}Fe_{0.2}Co_{0.2}Ni_{0.2})O_3$ are stabilized across a range of strain states. These films are shown to possess exceptional crystallinity, smooth film surfaces, and uniform mixing of the *3d* transition metal cations throughout the B-site sublattice as demonstrated by X-ray diffraction, atomic force microscopy, and scanning transmission electron microscopy. Field-dependent magnetometry shows that these materials have strong magnetic anisotropy and likely contain some amount of long-range magnetic order which can also be manipulated through lattice symmetry. This work suggests a new approach to design magnetic responses through designer combinations of B-site stoichiometries. The mechanisms driving functionality in these new high entropy oxide perovskites based on the *3d* transition metals are likely to be extremely complicated and may rely on an extraordinarily complicated exchange energy landscape, local structural-distortion-driven crystal field effects, or previously unobserved disorder-driven physical responses.

# ACKNOWLEDGEMENTS

Experiment design, sample synthesis, and structural characterization were supported by the US Department of Energy (DOE), Office of Basic Energy Sciences (BES), Materials Sciences and Engineering Division. A portion of this research was supported by the National Science Foundation under Grant No. DGE-1069091, Oak Ridge National Lab's Graduate Opportunities! Program, and the Department of Energy (DOE) Office of Science Graduate Student Research. STEM and some magnetometry was conducted through user proposal at the Center for Nanophase Materials Sciences, which is a US DOE, Office of Science User Facility.